\documentclass[preprint2,longabstract]{aastex}

\newcommand{\Mj}{$M_{\rm Jup}$}
\newcommand{\kms}{km\,s$^{-1}$}

\slugcomment{Revised version. Submitted to ApJ}

\shorttitle{Space velocities of L- and T-type dwarfs}
\shortauthors{Zapatero Osorio et al.}

\begin{document}

\title{Space Velocities of L- and T-type Dwarfs}

\author{M. R. Zapatero Osorio, E. L. Mart\'\i n\altaffilmark{1} and 
        V. J. S. B\'ejar}
\affil{Instituto de Astrof\'\i sica de Canarias, E-38205 La Laguna, 
       Tenerife, Spain}
\email{mosorio,ege,vbejar@iac.es}

\author{H. Bouy\altaffilmark{2}}
\affil{University of California at Berkeley, Astronomy Dept., 601 
       Campbell Hall, Berkeley CA 94720, USA}
\email{hbouy@astro.berkele.edu}

\author{R. Deshpande}
\affil{University of Central Florida, Department of Physics, P.O. Box 
       162385, Orlando, FL 32816-2385, USA}
\email{rohit@physics.ucf.edu}

\and

\author{R. J. Wainscoat}
\affil{Institute for Astronomy, 2680 Woodlawn Drive, Honolulu, HI 96822, 
       USA}
\email{rjw@ifa.hawaii.edu}

\altaffiltext{1}{Also at University of Central Florida, Dept$.$ of Physics, 
       P.\,O$.$ 162385, Orlando FL 32816-2385, USA}
\altaffiltext{1}{Also at Instituto de Astrof\'\i sica de Canarias, E-38205 
       La Laguna, Tenerife, Spain}

\begin{abstract}
We have obtained radial velocities of a sample of 18 ultracool dwarfs
with spectral types in the interval M6.5--T8 using high-resolution,
near-infrared spectra obtained with NIRSPEC and the Keck II
telescope. Among our targets there are two likely field stars of type
late M, one M6.5 Pleiades brown dwarf, and fifteen L and T likely
brown dwarfs of the solar neighborhood with estimated masses in the
range 30--75\,\Mj. Two dwarfs, vB\,10 (M8V) and Gl\,570\,D
(T7.5V/T8V), are known wide companions to low-mass stars. We have
confirmed that the radial velocity of Gl\,570\,D is coincident with
that of the K-type primary star Gl\,570\,A, thus providing additional
support for their true companionship. The presence of planetary-mass
companions around 2MASS\,J05591914$-$1404488 (T4.5V) has been analyzed
using five NIRSPEC radial velocity measurements obtained over a period
of 4.37\,yr. Using our radial velocity data and the radial velocities
of an additional set of eight L-type dwarfs compiled from the
literature in combination with their proper motions and trigonometric
parallaxes, we have computed $UVW$ space motions for the complete
expanded sample, which comprises a total of 21 L and T dwarfs within
20\,pc of the Sun. This ultracool dwarf population shows $UVW$
velocities that nicely overlap the typical kinematics of solar to
M-type stars within the same spatial volume. However, the mean
Galactic ($v_{\rm tot}$\,=\,44.2\,\kms) and tangential ($v_{\rm
t}$\,=\,36.5\,\kms) velocities of the L and T dwarfs appear to be
smaller than those of G to M stars. A significant fraction
($\sim$40\%) of the L and T dwarfs lies near the Hyades moving group
(0.4--2\,Gyr), which contrasts with the 10--12\%~found for
earlier-type stellar neighbors. Additionally, the distributions of all
three $UVW$ components ($\sigma_{UVW}$\,=\,30.2, 16.5, 15.8\,\kms) and
the distributions of the total Galactic ($\sigma_{v_{\rm
tot}}$\,=\,19.1\,\kms) and tangential ($\sigma_{v_{\rm
t}}$\,=\,17.6\,\kms) velocities derived for the L and T dwarf sample
are narrower than those measured for nearby G, K, and M-type stars,
but similar to the dispersions obtained for F stars. This suggests
that, in the solar neighborhood, the L- and T-type ultracool dwarfs in
our sample (including brown dwarfs) is kinematically younger than
solar-type to early M stars with likely ages in the interval
0.5--4\,Gyr.
\end{abstract}


\section{Introduction}

L- and T-type dwarfs are very cool ($T_{\rm eff}$\,$\le$\,2200\,K),
intrinsically faint objects of recent discovery
\citep{nakajima95,ruiz97,delfosse97,martin97,kirk05}, which are
frequently referred to as ``ultracool'' dwarfs. Over 500 of these
ultracool objects have been identified to date. Many of them are found
isolated in the solar neighborhood, i.e., within 50\,pc of the Sun;
only a few appear as widely separated ($a$\,$\ge$\,10\,AU) companions
to stars
\citep{nakajima95,rebolo98,burgasser00,kirk01,liu02,potter02,metchev04}
and as tight stellar companions \citep{freed03}. Yet their physical
and kinematic properties are not fully known. According to theoretical
evolutionary models \citep{burrows97,chabrier00}, the late L and T
dwarfs of the solar neighborhood are brown dwarfs with likely
substellar masses in the range 0.03--0.075\,$M_{\odot}$.

The detailed studies of low-resolution spectra and color--magnitude
diagrams of L and Tdwarfs
\citep{leggett02,patten06,kirk99,martin99,geballe02,burgasser02} are
beginning to shed light on the physical characteristics of these
low-mass objects because an increasing number of trigonometric
parallaxes (distances) have been determined for field ultracool dwarfs
\citep{dahn02,tinney03,vrba04,knapp04}. Besides distance and proper
motion, the additional ingredient for a complete kinematic study
employing space velocity components is radial
velocity. \citet{reid02}, \citet{mohanty03} and \citet{bailer04} have
obtained the first radial velocity measurements available for L-type
dwarfs in the solar vicinity using high-resolution optical
spectra. The marked low luminosities of the T dwarfs prevent 
accurate velocity determination at visible wavelengths.

Therefore, interpretation of the kinematics of the least massive
(including the substellar) population of the solar vicinity is less
developed than in the case of stars. Various groups have investigated
the dynamics and age of the Galactic disk by examining the observed
and simulated kinematics of nearby stars \citep{nordstrom04}. It is
found that the space velocity dispersion of stars increases with time
to the $\alpha$ power ($t^\alpha$) with $\alpha$\,=\,0.33
\citep{binney00}. Statistically, most solar-type to early M stars
within 50\,pc of the Sun show kinematic ages very similar to the age
of our solar system at about 5\,Gyr. A small fraction of these stars
turn out to be members of stellar moving groups characterized by much
younger ages from a few hundred megayears to a few gigayears
\citep{zuckerman04}. However, the nearby, coolest M dwarfs appear to
behave differently. As noted by \citet{reid02} and \citet{dahn02},
their kinematics suggest that the latest M stars of the solar
neighborhood are on average younger than earlier-type stars.

Here, we report on radial velocity measurements of ultracool L and T
dwarfs obtained from near-infrared spectra (Section 2). Radial
velocity variability has been investigated for a few targets in
Section 3. These data have been combined with trigonometric parallaxes
and proper motions published in the literature to derive galactic
velocities in Section 4. Our final remarks and conclusions are given
in Section 5.

\section{The sample and observations}

The sample of targets is listed in Table~\ref{vh}. It comprises three
late M-type dwarfs, one of which (PPl\,1 or Roque\,15) is a young,
lithium-bearing brown dwarf of the Pleiades cluster
\citep{stauffer98}. The remaining targets are six L- and nine T-type
field dwarfs recently discovered by the 2MASS, SLOAN, and Denis
surveys. The discovery papers are all indicated in Table~1 of
\citet{osorio06}. The spectral types given in the second column of
Table~\ref{vh} are taken from the literature
\citep{kirk97,kirk99,kirk00,martin99,geballe02,burgasser06,phan-bao06}. They
were derived from optical and/or near-infrared low-resolution spectra
and are in the interval M6.5--T8. There are objects that have been
classified differently by the various groups; we provide all
classifications in Table~\ref{vh}, first that of \citet{kirk99} and
\citet{burgasser06}, second that of \citet{geballe02}, and finally,
the classification from \citet{martin99}. In terms of effective
temperature, our sample spans the range 2700--770\,K
\citep{leggett00,vrba04}. J0334$-$49 and vB\,10 can be stars in our
sample while the remaining targets are very likely substellar.

We collected high-resolution near-infrared spectra of the 18 ultracool
dwarfs using the Keck II telescope and the NIRSPEC instrument, a
cross-dispersed, cryogenic echelle spectrometer employing a
1024\,$\times$\,1024 ALADDIN InSb array detector. These observations
were carried out on different occasions from 2000 December through
2006 January and are part of our large program aimed at the study of
radial velocity variability. The complete journal of the observations
is shown in Table~1 of \citet{osorio06}, since these data were
previously used to determine the rotational velocities of the sample,
including the young, lithium-bearing field brown dwarf
LP\,944$-$20. We note that the radial velocity of LP\,944$-$20 using
NIRSPEC spectra is fully discussed in \citet{martin06}. In the echelle
mode, we selected the NIRSPEC-3 ($J$-band) filter and an entrance
slit width of 0\farcs432 (i.e., 3 pixels along the dispersion
direction of the detector), except for eight targets (J2224$-$01,
J1728$+$39, J1632$+$19, J1346$-$00, J1624$+$00, J1553$+$15,
J1217$-$03, and GL\,570D) for which we used an entrance slit width of
0\farcs576. The length of both slits was 12\arcsec. All observations
were performed at an echelle angle of $\sim$63\arcdeg. This
instrumental setup provided a wavelength coverage from 1.148 up to
1.346\,$\mu$m split into ten different orders, a nominal dispersion
ranging from 0.164 (blue wavelengths) to 0.191\,\AA/pix (red
wavelengths), and a final resolution element of 0.55--0.70\,\AA~at
1.2485\,$\mu$m (roughly the central wavelength of the spectra),
corresponding to a resolving power
$R$\,$\sim$\,17800--22700. Individual exposure times were a function of
the brightness of the targets, ranging from 120 to 900\,s.

\subsection{Data reduction}

Raw data were reduced using the ECHELLE package within
IRAF\footnote{IRAF is distributed by National Optical Astronomy
Observatory, which is operated by the Association of Universities for
Research in Astronomy, Inc., under contract with the National Science
Foundation.}. Spectra were collected at two or three different
positions along the entrance slit. Nodded images were subtracted to
remove sky background and dark current. White-light spectra obtained
with the same instrumental configuration and for each target were used
for flat-fielding the data. All spectra were calibrated in wavelength
using the internal arc lamp lines of Ar, Kr, and Xe, which were
typically acquired after observing the targets. The vacuum wavelengths
of the arc lines were identified and we produced fits using a 
third-order Legendre polynomial along the dispersion axis and a second-order
one perpendicular to it. The mean rms of the fits was 0.03\,\AA, or
0.7~\kms. In order to correct for atmospheric telluric absorptions,
near-infrared featureless stars of spectral types A0--A2 were observed
several times and at different air masses during the various observing
runs. The hydrogen line at 1.282\,$\mu$m, which is intrinsic to these
hot stars, was removed from the spectra before using them for division
into the corresponding science data. Finally, we multiplied the
science spectra by the black-body spectrum for the temperature of
9480\,K, which is suitable for A0V type \citep{allen00}.

\begin{figure}
\epsscale{1.0}
\plottwo{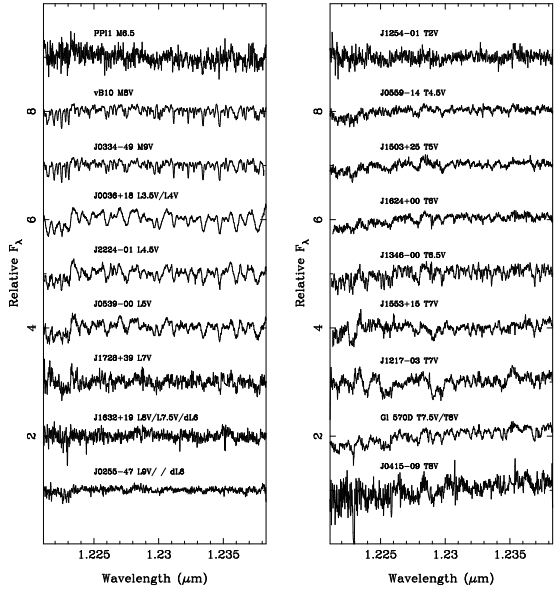}{f1b.ps}
\caption{NIRSPEC spectra of our sample centered at 1.23~$\mu$m. All
spectra are normalized to unity, offset by 1 on the vertical axis, and
shifted in velocity to vacuum wavelengths.
  \label{spec_o05}}
\end{figure}

\begin{figure}
\epsscale{1.0}
\plottwo{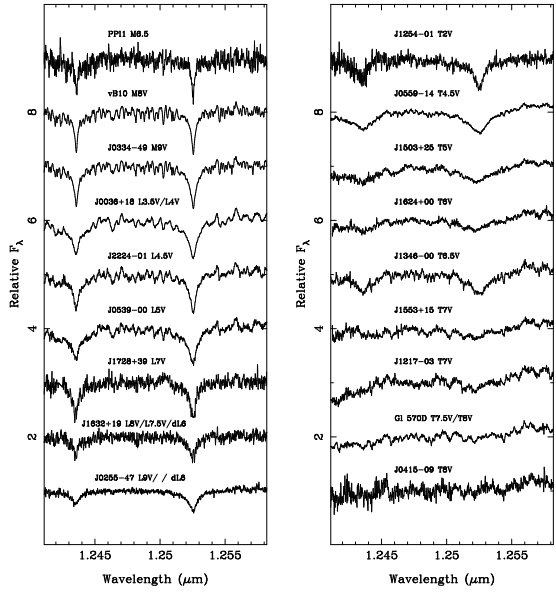}{f2b.ps}
\caption{NIRSPEC spectra of our sample centered at 1.25~$\mu$m. The
most prominent features of the M and L dwarfs are due to K\,{\sc i}
absorption, which vanishes at late T types. All spectra are offset by
1 on the vertical axis, and shifted in velocity to vacuum
wavelengths. Data are normalized to unity in the intervals
1.2465--1.2475~$\mu$m (M6.5--T4.5) and 1.2468--1.2472~$\mu$m (T5--T8).
  \label{spec_o06}}
\end{figure}

\begin{figure}
\epsscale{1.0}
\plottwo{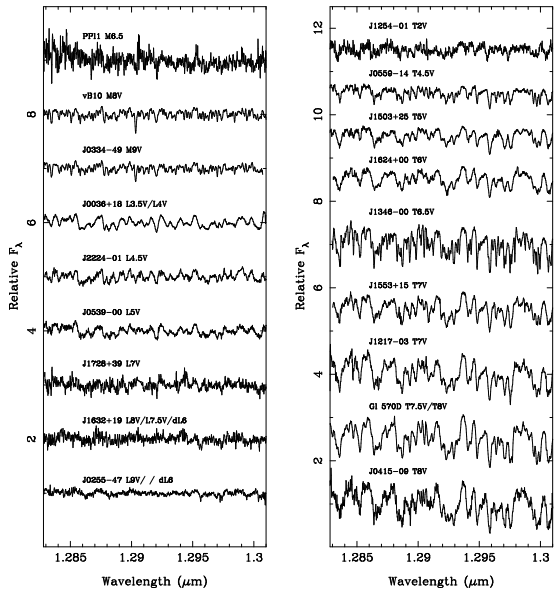}{f3b.ps}
\caption{NIRSPEC spectra of our sample centered at 1.292~$\mu$m. All
spectra are normalized to unity and shifted in velocity to vacuum
wavelengths. The spectra of the left panel are offset by 1 on the
vertical axis. Data in the right panel are offset by 1 (T2V--T6V
dwarfs, top four spectra) and by 1.5 (T6.5V--T8V dwarfs, bottom five
spectra).
  \label{spec_o08}}
\end{figure}

We have plotted in Figures~\ref{spec_o05} to~\ref{spec_o08} all of our
spectra corresponding to the echelle orders centered at 1.230\,$\mu$m,
the K\,{\sc i} doublet (1.2485\,$\mu$m), and 1.292\,$\mu$m. These are
orders relatively free of strong telluric lines. Effective temperature
(i.e., spectral type) decreases from top to bottom. Note that all
spectra have been shifted in velocity to vacuum wavelengths for easy
comparison of the atomic and molecular features. The signal-to-noise
ratio of the data varies for different echelle orders and different
targets depending on their brightness. In general, the red orders show
better signal-to-noise ratio than the blue orders, except for the
reddest order centered at 1.337\,$\mu$m in T dwarfs since these
wavelengths are affected by strong methane and water vapor absorptions
below 1300\,K.

\subsection{Data analysis: radial velocities}

We note that systematic errors or different zero-point shifts may be
present in the instrumental wavelength calibration of our data, which
may affect the radial velocity measurements. Various authors
\citep{tinney98,martin99b,mohanty03,basri_reiners06} have reported on
the stable heliocentric velocity of the M8V dwarf vB\,10, finding an
average value of $+$35.0\,\kms. From our two NIRSPEC spectra and using
the centroids of the K\,{\sc i} lines, we derived $+$35.0 and
$+$34.3\,\kms~with an estimated uncertainty of
$\pm$1.5\,\kms~associated with each individual determination. These
measurements are fully consistent with those in the
literature. Furthermore, we have compared NIRSPEC spectra
corresponding to echelle orders that contain a considerable number of
telluric lines and that were observed during a night and on different
nights. We found that the telluric lines typically differ by less than
1\,\kms~in velocity. All this suggests that any systematic errors or
different zero-point shifts in our radial velocities are likely
smaller than the measurement uncertainties, which are typically
$\ge$1\,\kms.

\begin{figure}
\epsscale{1.0}
\plottwo{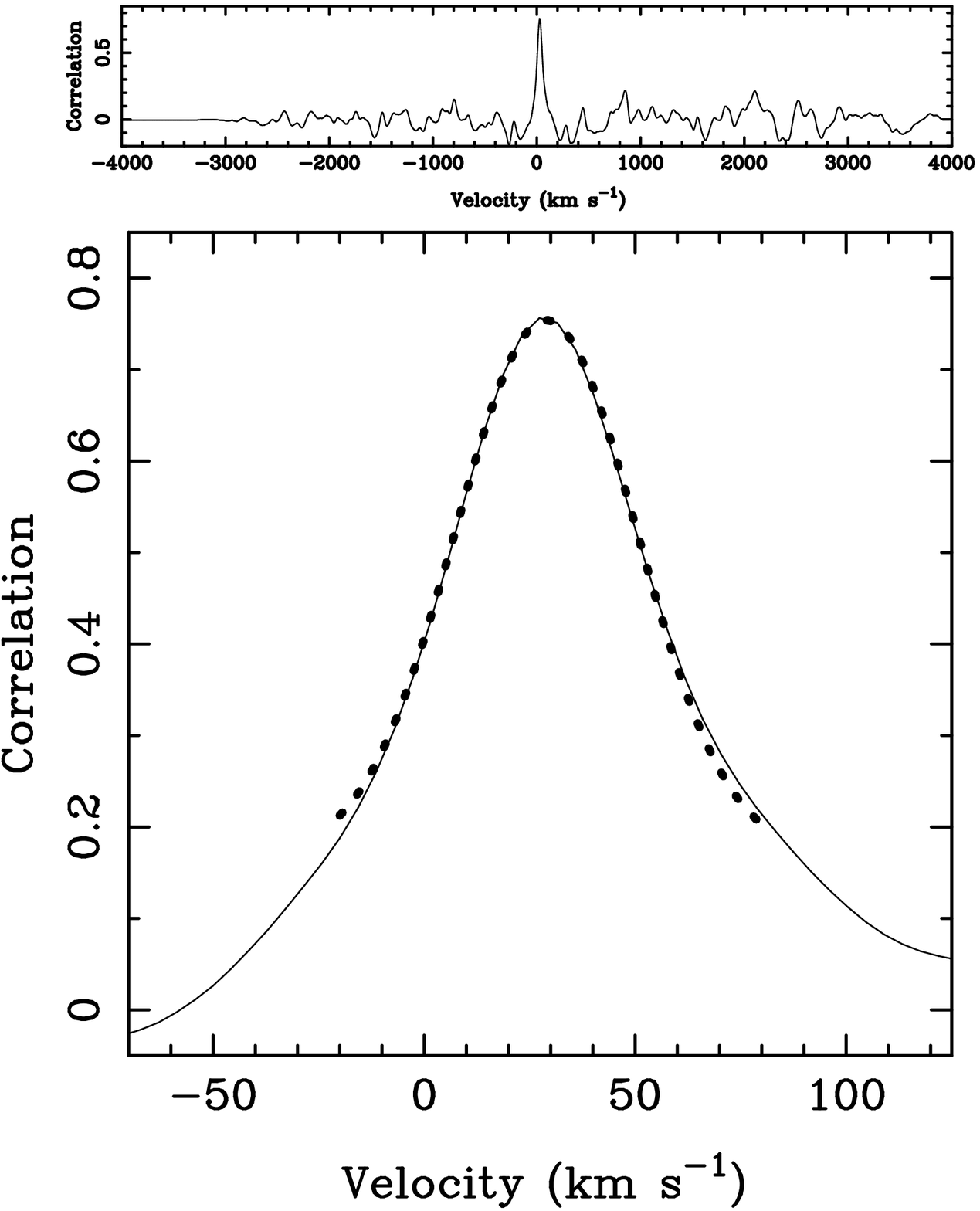}{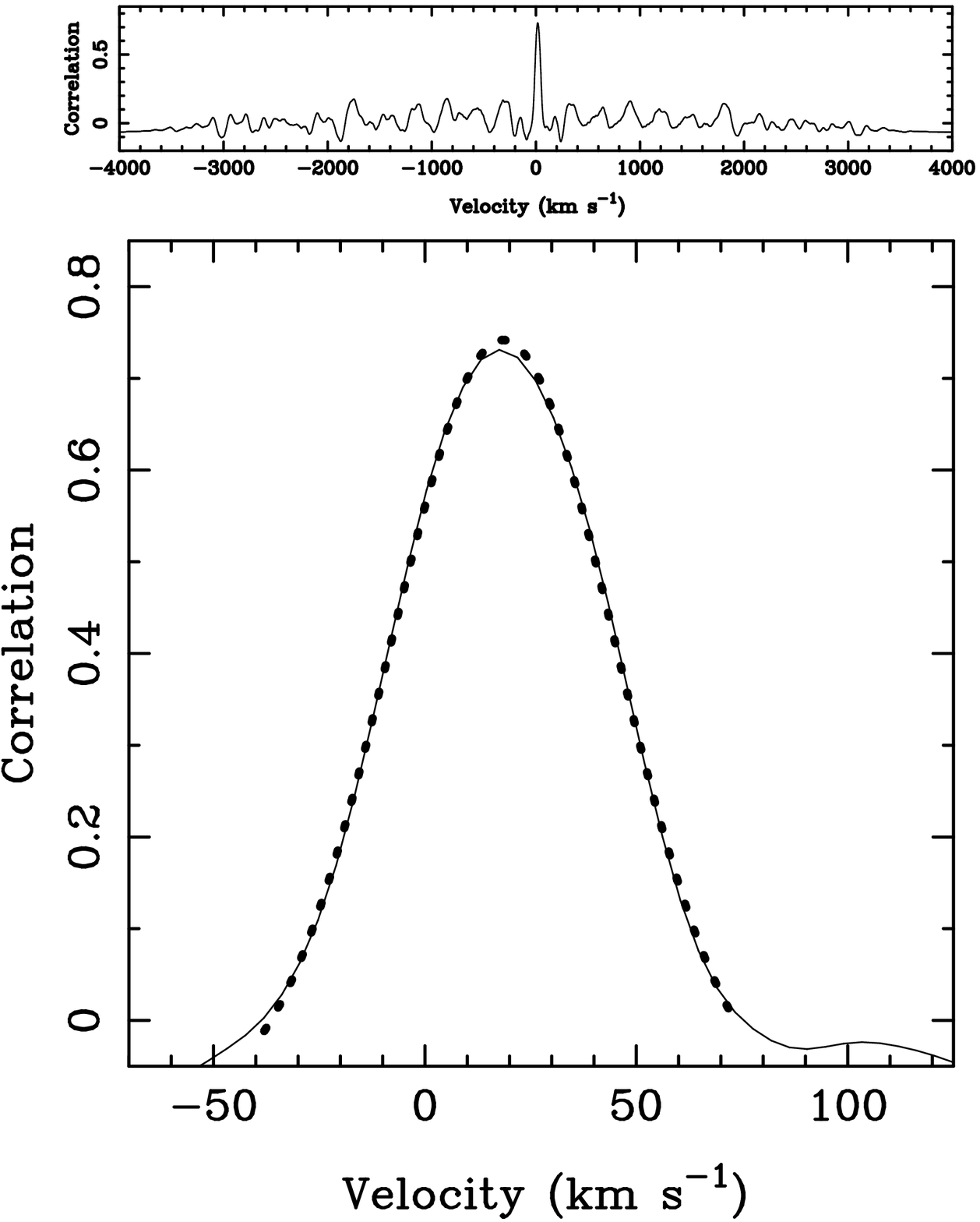}
\caption{Cross-correlation functions of Gl\,570\,D against J0559$-$14
(left panel, spectra centered at 1.292\,$\mu$m), and J0036$+$18 against
vB\,10 (right panel, spectra centered at 1.230\,$\mu$m). The bottom
panels show an enlargement around the maximum peak of the
cross-correlation functions (thin line) and the best Gaussian fits to
the core of the peak (thick dotted line).
  \label{gauss}}
\end{figure}

We derived heliocentric radial velocities, $v_h$, via a
cross-correlation of the spectra of our tarets against spectra of
dwarfs of similar types with known heliocentric velocity. The details
of the procedure are fully described in the literature
\citep{marcy89}. Summarizing, heliocentric radial velocities were
obtained from the Doppler shift of the peak of the cross-correlation
function between the targets and the templates. To determine the
center of the cross-correlation peak we typically fit it with a
Gaussian function using the task FXCOR in IRAF. This provides observed
velocities that are corrected for the Earth's motions during the
observations and converted into heliocentric velocities. Only orders
for which we unambiguously identified the peak of the
cross-correlation function and obtained good Gaussian fits were
employed (see Figure~\ref{gauss} for examples of cross-correlation
functions and Gaussian fits). We typically used between 4 and 8
echelle orders depending on the signal-to-noise ratio of the data. All
results were then averaged to produce the final
$v_h$~measurements. Uncertainties are derived from the standard
deviation of the mean.

We used the M8V-type dwarf vB\,10 as the primary reference/template
object for several reasons: first, it is a slow rotator
\citep{mohanty03}, thus providing narrow cross-correlation peaks;
second, its radial velocity is well determined in the literature to be
$+$35.0\,\kms; and third, the signal-to-noise ratio of its NIRSPEC
spectra is reasonably good (see Figures~\ref{spec_o05}
to~\ref{spec_o08}), minimizing the data noise introduced in the
cross-correlation method. This technique assumes that the target and
template spectra are of similar type and differ only in the rotation
velocity. Nevertheless, \citet{bailer04} has recently shown that
M-type templates can also yield accurate velocities for L dwarfs. In
our sample, the energy distribution of T dwarfs does differ
significantly from M dwarfs, and we found that for the early T-type
targets the cross-correlation with vB\,10 gives reasonable results if
the resonance lines of K\,{\sc i} are avoided and only molecular
(particularly water vapor) lines are used in the
cross-correlation. For the late-type objects, cooler templates are
required. We employed J2224$-$01 (L4.5V), J0559$-$14 (T4.5V), and
J1217$-$03 (T7--8V) as secondary reference dwarfs to obtain the radial
velocities of the L- and T-type objects in the sample. Rotation
velocities of all targets are provided in column~10 of
Table~\ref{vh}~\citep{osorio06}. The three secondary templates have
moderate rotation in the interval $v_{\rm
rot}$\,sin\,$i$\,=\,22--31\,\kms. However, the derived error bars on
radial velocity do not appear to be significantly higher than using
the slow rotator vB\,10, suggesting that, for the spectral resolution
and signal-to-noise ratio of our data, radial velocity accuracy shows
little dependence on rotation velocity up to $v_{\rm rot}$\,sin\,$i$
$\sim$ 30\,\kms.

Columns 5 through 8 of Table~\ref{vh} show our measured $v_h$ for all
objects observed on different occasions in this program. We also give
the observing dates in columns 3 (Universal Time) and 4 (Modified
Julian Date). Whenever there is more than one observation available
per object, the first spectrum also acts as a template spectrum to
obtain radial velocities via the cross-correlation technique, and the
derived velocities are provided in column 9 of Table~\ref{vh}. This
minimizes the effects of using templates of different spectral types
and serves as a test of consistency. Furthermore, such a procedure will
allow us to study binarity. Previous radial velocities reported in the
literature are listed in the eleventh column of the Table. We note
that, with the exception of PPl\,1 (see below), all of our measurements
agree with the literature values to within 1-$\sigma$ of the claimed
uncertainty.

\section{Binarity}

Gl\,570D is a known wide companion ($\sim$1500\,AU) to the multiple
system Gl\,570ABC \citep{burgasser00}. We have determined its
heliocentric velocity to be $+$28.5\,$\pm$\,2.7\,\kms, which is
coincident with the radial velocity of the K-type star Gl\,570A
($v_h$\,=\,$+$27.0\,$\pm$\,0.3\,\kms) as measured by
\citet{nidever02}. This suggests that Gl\,570D is not a short-period
binary of mass ratio near 1. The secondary companion, Gl\,570BC, is a
spectroscopic binary whose radial velocities span the range
8.8--37.9\,\kms~\citep{marcy89}.

Multiple NIRSPEC spectra obtained on different occasions are available
for five dwarfs in our sample. These are shown in Table~\ref{meanv},
where we also provide the time elapsed between the first and last
observation and the mean heliocentric velocity per object with the
associated average error bar. Inspection of all velocities listed in
Table~\ref{vh} reveals no obvious trace of moderate velocity
perturbation in any of the dwarfs since all their measured $v_h$ 
agree within the uncertainties over the time span of the data. This
indicates that the dwarfs of Table~\ref{meanv} are not very likely to be 
close equal-mass binaries.

\begin{figure}
\epsscale{1.0}
\plotone{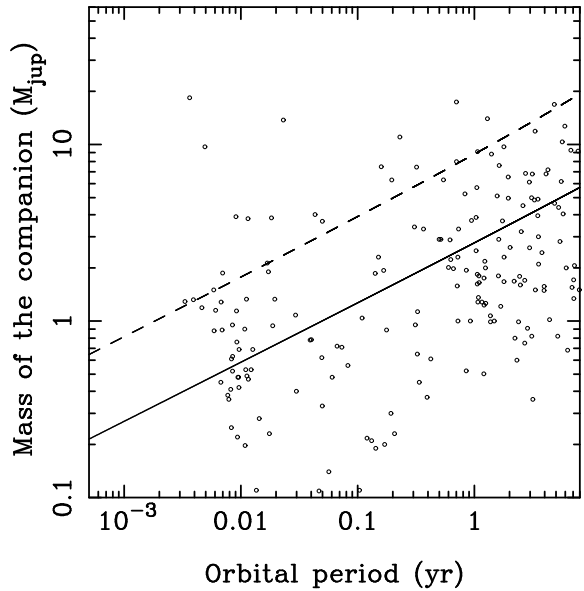}
\caption{Detectability of substellar companions around the T4.5V dwarf
J0559$-$14 using our NIRSPEC data. Companions with masses above the
curves can be excluded with a confidence of 1-$\sigma$ (solid line)
and 3-$\sigma$ (dashed line). As indicated in the text, the time
coverage of our data allows us to study orbits with periods of a few
days and a few years; orbits in between are not well sampled. To
compute the detectability curves we have adopted circular orbits
parallel to the line of sight, a mass of 0.06\,$M_{\odot}$ and a
radial velocity amplitude of 0.5\,\kms~for the primary
component. Known radial velocity planets around stars are plotted as
open circles.
  \label{j0559}}
\end{figure}

To relate companion mass to the dispersion of our radial velocities we
use the mass function derived from the Keplerian equations
\citep{marcy89}. The most constraining case in our sample is the T4.5V
dwarf J0559$-$14. For it we can study the presence of planetary-mass
companions in short (a few days or $\sim$0.01~AU) and intermediate (a
few years or $\sim$0.4~AU) orbits. J0559$-$14 radial velocity curve
exhibits 1-$\sigma$ standard deviation of 0.50\,\kms. For a primary
mass of about 60\,\Mj, expected for mid-T field dwarfs with an age of
$\sim$5\,Gyr \citep{chabrier00}, and at the 3-$\sigma$ level,
companions more massive than 2\,\Mj~(short periods) and
10\,\Mj~($\sim$1-yr period) can be excluded. Figure~\ref{j0559}
summarizes these results graphically. The region of parameter space
excluded for planets around J0559$-$14 is comparable to the region
excluded around stars. However, because of the time coverage of our
observations of J0559$-$14, we cannot reach firm conclusions on the
presence of companions with orbital periods between days and less than
a year. Circular orbits have been adopted to compute these
estimates. We consider these minimum mass estimates to represent
approximate guidelines in future corroborative spectroscopic work.

Regarding J0036$+$18 and J0539$-$00, the null detection of radial
velocity variations over a few years suggests that no companions of
$\sim$10\,\Mj~or more can exist near them. Similarly, very close-in
companions of a few Jupiter masses may be ruled out orbiting PPl\,1 or
J0334$-$49 with a periodicity of a few days. However, we stress
that two radial velocity data points constitute statistically too few
observations to reach a final conclusion on the presence of Jovian
planets around these objects. Further measurements are certainly
required to constrain firmly the minimum mass of any possible
companion. However, this exercise indicates that companions with
masses in the planetary domain lie near the detectability limit in our
data.

Interestingly, \citet{martin98} measured the heliocentric velocity of
the Pleiades brown dwarf PPl\,1 at $+$15.4\,$\pm$\,1.6\,\kms, which
deviates from our measurement (see Table~\ref{meanv}) by more than
4\,$\sigma$. Despite the fact that such a difference may suggest the
presence of a second object (future observations are pending to test
this hypothesis), its total amount seems to be rather large for a
planetary companion even though it might account for the peak-to-peak
amplitude of the primary radial velocity curve. This may indicate that
the companion has a broad orbit and a mass above the deuterium-burning
mass limit, i.e., $\ge$13\,\Mj, thus, it belongs to the brown dwarf
regime. From the locus of PPl\,1 in the $K$ vs $(I-K)$ color-magnitude
diagram of the Pleiades cluster, \citet{pinfield03} concluded that
this object is a probable binary with a mass ratio in the interval
$q$\,=\,0.75--1. We note that our NIRSPEC heliocentric velocity of
PPl\,1 is fully consistent with the systemic velocity of the Pleiades
measured at $+$5.4\,$\pm$\,0.4\,\kms~\citep{liu91,kharchenko05}, hence
providing additional evidence for its membership of this star cluster.

\section{Galactic space motions and kinematics}

Using our radial velocity measurements and the proper motions and
trigonometric parallaxes provided in the literature by \citet{dahn02},
\citet{vrba04}, \citet{knapp04}, and \citet{an07} we have calculated
the $U$, $V$, and $W$ heliocentric velocity components in the
directions of the Galactic center, Galactic rotation and north
galactic pole, respectively, with the formulation developed by
\citet{johnson87}. Note that the right-handed system is employed and
that we will not subtract the solar motion from our calculations. The
uncertainties associated with each space-velocity component are
obtained from the observational quantities and their error bars after
the prescriptions of \citet{johnson87}. Our derivations are shown in
Table~\ref{vgal}. Unfortunately, two T dwarfs in the sample lack
distance and proper motion measurements, thus preventing us from
determining their Galactic velocities (these dwarfs are not shown in
Table~\ref{vgal}). No trigonometric parallaxes are available for
PPl\,1 and the M9V dwarf J0334$-$49; we have adopted the Pleiades
distance for the former object \citep{an07} and the poorly constrained
spectroscopic distance estimate of \citet{phan-bao06} for the latter,
which result in relatively large uncertainties associated with the
derived $UVW$ velocities.

With the only exceptions of PPl\,1 and J1728$+$39AB, the remaining
objects in our sample with known trigonometric parallax lie within
20\,pc of the Sun. To carry out a reliable statistical study of the
kinematics of the least massive population in the solar neighborhood,
we have enlarged our sample of ultracool dwarfs with an additional set
of eight L-type objects of known trigonometric distance
($d$\,$\le$\,20\,pc), proper motion and radial velocity
\citep{dahn02,mohanty03,bailer04,vrba04}. This additional set is shown
in Table~\ref{vgal_literature}, where we also provide our $UVW$
computations obtained in the manner previously
described. Hence, $UVW$ space velocities are finally available for
a total of 21 L and T dwarfs located at less than $\sim$20\,pc from
the Sun.

\subsection{Star moving groups}

\begin{figure}
\epsscale{1.0}
\plottwo{f6aa.ps}{f6ab.ps}
\plottwo{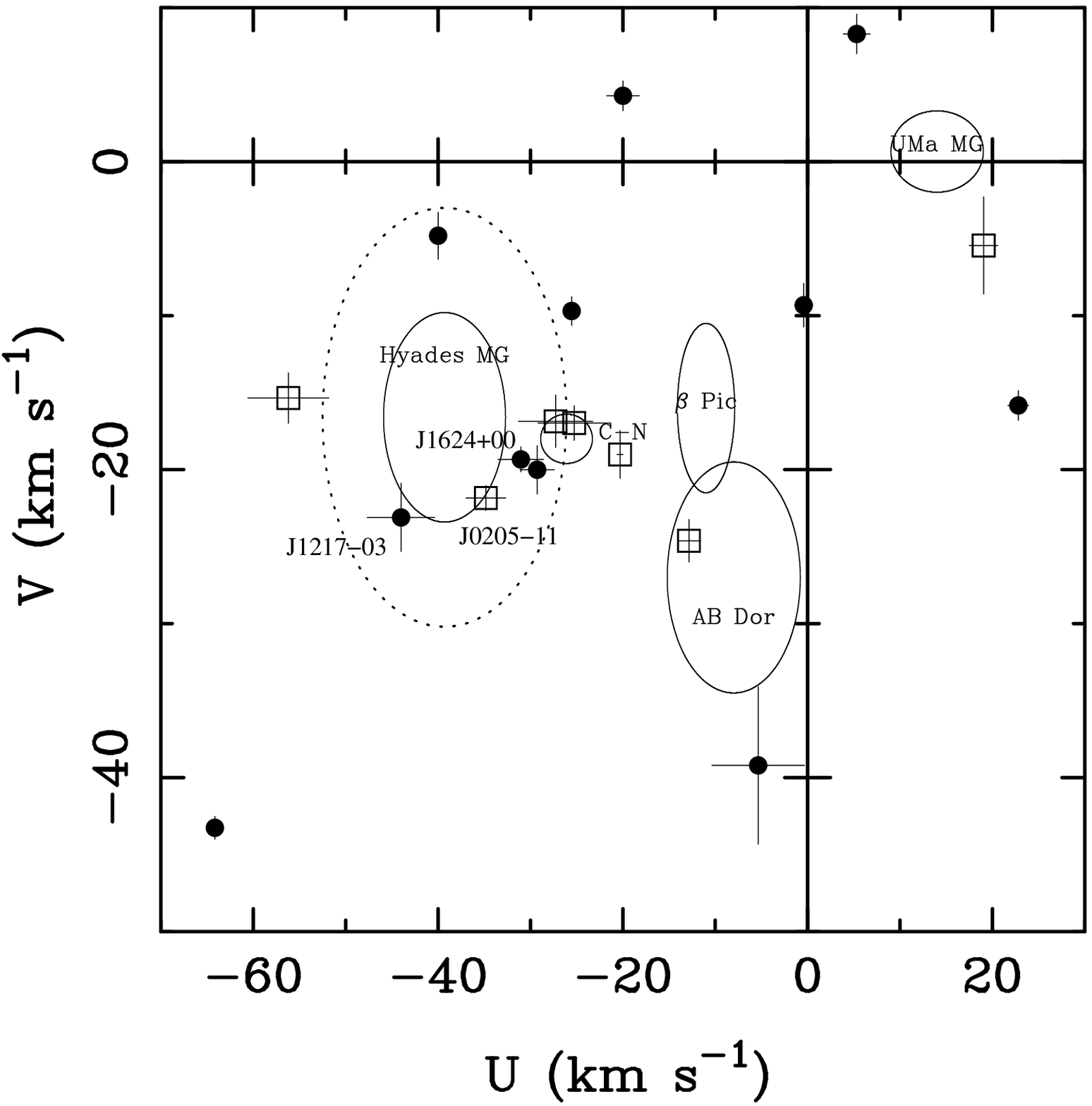}{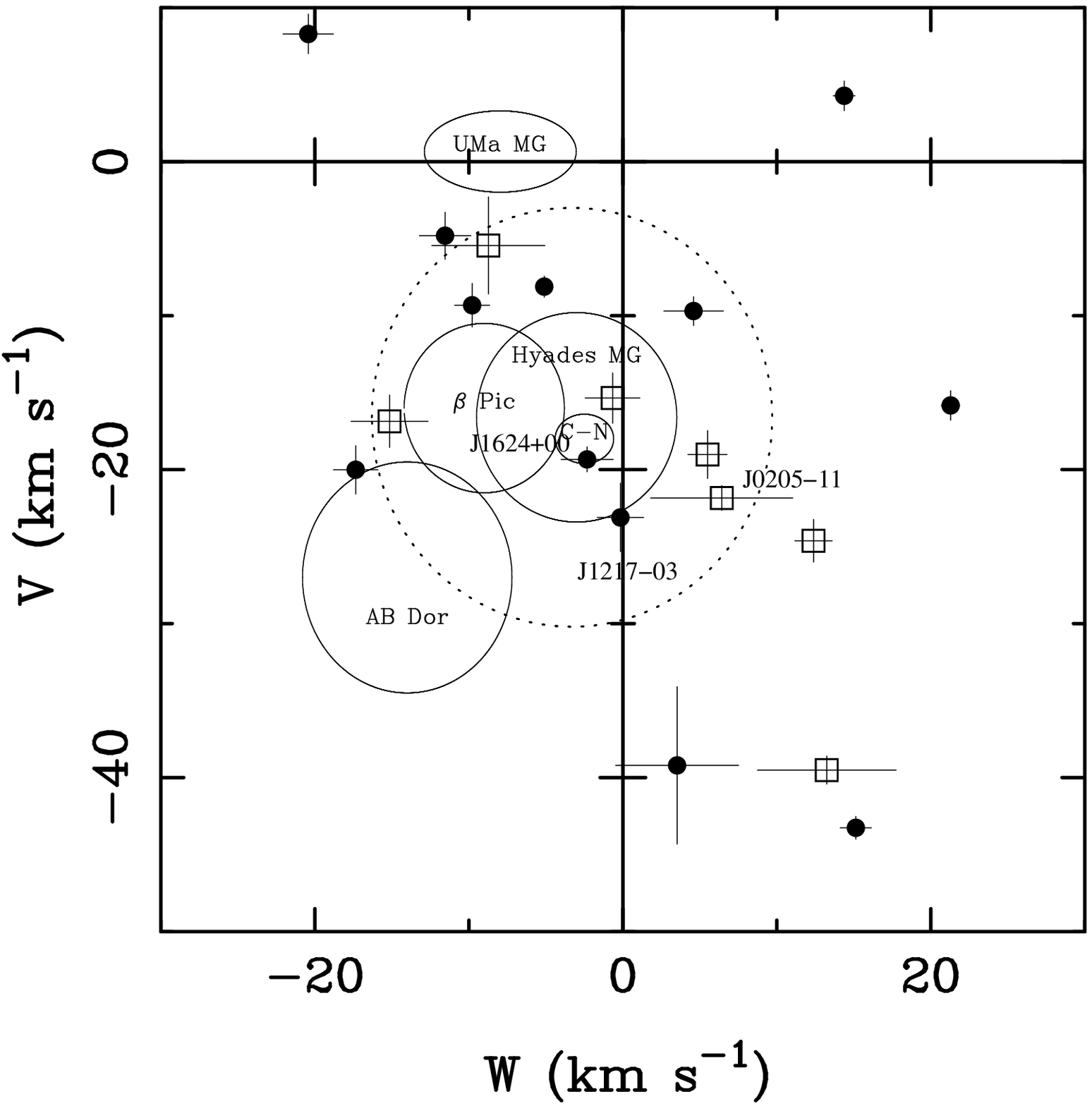}
\caption{Space motions of our sample of ultracool dwarfs (filled
  circles) and the L dwarfs from the literature (open squares) with
  $d$\,$\le$\,20\,pc. The young Pleiades brown dwarf PPl\,1 is
  indicated in the top panels. An enlargement is shown in the bottom
  panels along with the position of various nearby star moving groups
  (ellipses, C-N stands for Carina-near). Solid line ellipses stand
  for the 1-$\sigma$ location of the moving groups, and the dotted
  ellipse represents the 2-$\sigma$ width of the Hyades moving
  group. The Pleiad PPl\,1 is not included in the bottom panels.
  \label{uvw_planes}}
\end{figure}

For the following discussion, we will focus on the extended sample of
L and T dwarfs. Their Galactic motions are depicted in the $UV$ and
$VW$ planes in Fig.~\ref{uvw_planes}. As can be seen from the top
panels, the space velocities of most L and T dwarfs appear to be
rather scattered in these diagrams, and no preferred region (or
clustering) of high density and small velocity dispersion is
perceptible. \citet{leggett92} summarized the criteria used for
kinematically classifying old/young disk and halo stars as follows:
objects with $V$\,$<$\,$-$100\,\kms~or high eccentricity in the $UV$
plane are defined to be halo stars; objects with an eccentricity in
the $UV$ plane of $\sim$0.5 are defined to be old disk-halo stars;
objects with $-$50\,$<$\,$U$\,$<$\,$+$20, $-$30\,$<$\,$V$\,$<$\,0, and
$-$25\,$<$\,$W$\,$<$\,$+$10~\kms~(i.e., the ``young disk'' ellipsoid)
are defined to be young disk sources. Stars that lie outside this
ellipsoid are defined to be old disk or young-old disk objects
depending on their $W$ velocity. Employing these criteria and from the
inspection of the $UVW$ velocities, none of the L and T dwarfs in our
study seems to belong to the old-disk or halo kinematic categories;
all of them can actually be grouped into the young disk and young--old
disk kinematic classifications, suggesting that they are likely
objects with ages typical of the solar system and younger.

To further develop this idea, we have superimposed in the bottom
panels of Fig.~\ref{uvw_planes} the location of known, nearby star
moving groups (streams of stars with common motion through the Milky
Way). The $UVW$ velocities and their associated 1-$\sigma$ velocity
dispersions used to depict the ellipsoids of the various moving groups
shown in Fig.~\ref{uvw_planes} are taken from the literature
\citep{eggen92,chen97,dehnen98,chereul99,montes01,zuckerman04,famaey05,zuckerman06}.
Ordered by increasing age, the $\beta$~Pictoris (10--30\,Myr), AB Dor
($\sim$50\,Myr), Carina-near ($\sim$200\,Myr), Ursa Majoris
($\sim$300\,Myr), and Hyades (0.4--2\,Gyr) moving groups have been
selected for the proximity of all their northern and southern stellar
members to the Sun (typically less than 55\,pc). The open star cluster
Hyades ($\sim$625~Myr) lies within the Hyades moving group. Other
young moving groups and star clusters of recent discovery are not
included in Fig.~\ref{uvw_planes} because they only contain southern
stars, like Tucana/Horologium and the TW Hydrae association, or are
located farther away, like $\eta$ Cha and the Pleiades moving group
\citep{zuckerman04}. The Hyades moving group is the oldest dynamical
stream considered in this work, with stellar ages (derived from A-F
stars) spanning the interval 400\,Myr to $\sim$2\,Gyr
\citep{chereul99}, but probably younger than the canonical age of the
solar system.

From the bottom $UV$- and $WV$-planes of Fig.~\ref{uvw_planes}, it
becomes apparent that only three dwarfs in our sample, namely
J1624$+$00 (T6V), J1217$-$03 (T7V), and J0205$-$11AB (L7V), all have
space velocity components roughly consistent with the Hyades moving
group and fall inside the 1.5-$\sigma$ ellipsoid. These are labeled in
the bottom panels of Figure~\ref{uvw_planes}. We note that
J0205$-$11AB is an equal-mass binary separated by 9.2\,AU
\citep{koerner99}, and that the expected amplitude of the radial
velocity curve of each component is therefore less than 3\,\kms~for
masses below the substellar limit. This has very little impact on the
derived $UVW$ velocities of J0205$-$11AB.

Our results for J1624$+$00, J1217$-$03, and J0205$-$11AB are in
agreement with the previous work by \citet{bannister07}. These authors
also found that J0036$+$18 (L3.5V/L4V) and J0825$+$21 (L7.5V) show
proper motion consistent with the Hyades moving group; our $UVW$
measurements lie close to the 2-$\sigma$
ellipsoid. \citet{bannister07} also commented on the fact that all
these five ultracool dwarfs sit on a very tight sequence in
color-magnitude diagrams (suggesting they are coeval objects)
consistent with evolutionary models of 500~Myr. Nevertheless, recent
works by \citet{famaey05} and \citet{simone04} caution against
assigning ages based solely on space motion and star moving group
memberships. On the contrary, membership in open star clusters
provides reliable age estimates. None of the L and T dwarfs in our
sample appears to be unambiguous members of the Hyades star
cluster. Recently, \citet{bihain06} found that the velocity (or proper
motion) dispersion of Pleiades brown dwarfs ($\sim$120\,Myr) is about
a factor of four times higher than that of solar-mass stars in the
cluster. In our sample, only J1217$-$03 has space velocities close to
the 4-$\sigma$ ellipsoid centered at $U$\,=\,$-$42, $V$\,=\,$-$19,
$W$\,=\,$-$1~\kms~\citep{bruijne01}, which corresponds to the
$\sim$625~Myr-old Hyades.

Further study of the bottom $UV$- and $WV$-planes of
Fig.~\ref{uvw_planes} reveals that about 40\%~of the L and T dwarfs
lie near the Hyades moving group either within the 2-$\sigma$
ellipsoid or ``touching'' it (i.e., the error bars cross the
2-$\sigma$ ellipsoid). To analyze the significance of such an apparent
concentration, we have collected accurate radial velocities, Hipparcos
distances and parallaxes of F-, G-, K- and M-type stars within 20\,pc
of the Sun from the catalog of \citet{kharchenko04}. A total of 10 F
stars, 25 solar-type stars, 58 K, and 52 early M (M0--M5) stars were
selected. Their Galactic $UVW$ motions were calculated as for our L--T
sample contained within the same spatial volume. The left panels of
Fig.~\ref{distributions} show the space velocities of these stars and
of our extended sample as a function of heliocentric distance.

After inspection of the space velocity distribution of the G to early
M stars in the $UV$- and $WV$-planes, we found that only 10--12\%~of
the stars (very similar statistics for all three stellar spectral
types) lie within or in the ``surroundings'' of the Hyades moving
group, which contrasts with the higher value, 40\%, derived for the
ultracool dwarfs. Furthermore, under the assumption that the L--T
population follows a Poissonian distribution with the same object
density than the G to early-M stars, we have estimated the probability
of finding eight out of 21 ($\sim$40\%) ultracool dwarfs close to the
Hyades moving group to be lower than 1\%, indicating that this
concentration is significant to a confidence level of about 99\%.
This suggests that, on average, the L and T population is
kinematically younger than the majority of the stars in the solar
neighborhood.

\subsection{Galactic kinematics and age of the ultracool dwarfs}

\begin{figure}
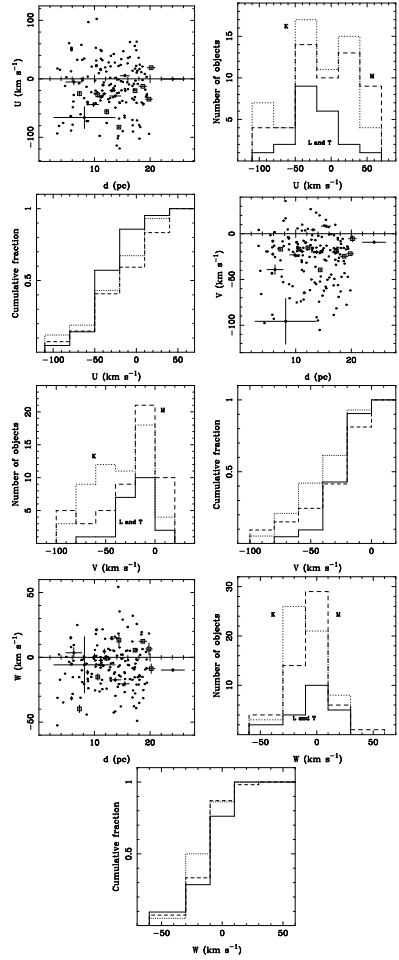

\epsscale{0.32}
\plotone{f7_1a.ps}
\plotone{f7_1b.ps}
\plotone{f7_1c.ps}
\plotone{f7_2a.ps}
\plotone{f7_2b.ps}
\plotone{f7_2c.ps}
\plotone{f7_3a.ps}
\plotone{f7_3b.ps}
\plotone{f7_3c.ps}
\caption{{\sl (Left panels).} Space velocities of our sample of
ultracool dwarfs (filled symbols), the extended sample (open squares),
and the G, K, and M0--M5 stars (tiny circles). Only the error bars of
the ultracool dwarfs are depicted for the clarity of the figure. {\sl
(Middle panels).} $UVW$ distributions for the various populations (K
stars --- dotted histogram; early-M stars --- dashed histogram; L and
T dwarfs --- solid histogram). {\sl (Right panels).} Cumulative
distributions of space velocities.
  \label{distributions}}
\end{figure}

From the left panels of Fig.~\ref{distributions} it is evident that
the $UVW$ components of the L and T dwarfs overlap the range of space
velocities observed for the G, K, and early M stars, i.e., they are
neither larger nor dramatically smaller despite the reduced mass of
the L- and T-type objects.  We can discuss this further by producing
the histograms shown in the middle panels of Fig.~\ref{distributions},
which depict the $UVW$ distributions for the three populations (K-type
stars, early M stars, and the L and T ultracool dwarfs). The
cumulative distributions are displayed in the right-most panels of
Fig.~\ref{distributions}. We can compare these distributions
quantitatively using the Kolmogorov--Smirnov test. Such comparison
globally shows that there is a probability of more than 15\%~that the
ultracool dwarfs and the K and M0--M5 stars are drawn from the same
kinematic population. While this holds for a comparison between the
$U$ and $W$ velocity dispersions of the L--T dwarfs and the K--M
stellar samples, the test suggests larger differences for the $V$
distributions (galactic rotation) of the K-type stars and the
ultracool dwarfs; however, these differences are not statistically
meaningful and we will not discuss them further.

In addition, the width of the histograms of Fig.~\ref{distributions},
represented by the parameters $\sigma_U, \sigma_V, \sigma_W$, is
relevant for a complete kinematic analysis. It is well established
that all three velocity dispersions increase with age
\citep{spitzer51,mayor74,wielen77}. We have obtained $\sigma_U,
\sigma_V$, and $\sigma_W$ for each stellar and substellar populations
(see Table~\ref{sigmas}). The Galactic motion dispersions of the G- to
M5-type stars are in full agreement with the values obtained for the
thin disk of the Galaxy available in the literature
\citep{hawley96,bensby03}, even though all these works include a
larger number of stars and volumes in their studies. All the L and T
dwarfs in our sample have typical perpendicular distances to the
Galactic plane in the interval $\pm$15\,pc. Our $\sigma_U$ and
$\sigma_W$ derivations for the F-type stars also coincide with the
literature; however, the derived $\sigma_V$ value appears smaller than
the 20\,\kms~found for the thin disk A and F stars by
\citet{bensby03}. This could be explained as statistical noise due to
the relative small number (10) of F stars in our study.

We have found that the Galactic velocity distributions corresponding
to the ultracool L--T dwarfs are narrower (i.e., lower dispersions in
all three space velocities) than those of G to M5 stars. However, our
derived $\sigma_U, \sigma_V$, and $\sigma_W$ for the L--T dwarfs are
remarkably similar to results obtained for M-type emission-line stars
\citep{hawley96} and very active M7-type stars \citep{west06}, which
are generally assigned ages of less than $\sim$3\,Gyr
\citep{reid02}. We also note that if the $\sim$40\%~of the L and T
dwarfs lying within and near the Hyades moving group in the $UVW$
planes is removed from the statistical analysis, the three parameters
$\sigma_U, \sigma_V$, and $\sigma_W$ will gently increase;
nevertheless, they remain smaller than the Galactic velocity
dispersions observed for solar to early-M stars. Interestingly, all
three $\sigma_{UVW}$ of L and T dwarfs resemble the velocity
dispersions of F-type stars (Table~\ref{sigmas}), suggesting that the
two populations (despite their very different mass) share similar ages
on average.

There are various empirical calibrations of the relationship between
velocity dispersion, particularly $\sigma_W$, and age
\citep{mayor74,haywood97,nordstrom04} that we can use to estimate the
mean age for our kinematic samples. According to the $\sigma$-age
relations provided in Tables~2 (A and F stars) and~3 (G-K-M stars) by
\citet{mayor74}, the mean age of the 20-pc sample of F stars is about
2\,Gyr, and that of the G to early M stars is about 5\,Gyr (i.e., the
age of the Sun). \citet{wielen77} published a mathematical relation
between age and the dispersion of the total Galactic velocity (see
also \citet{schmidt07}). We have computed the mean Galactic velocity,
$<v_{\rm tot}>$, and its associated dispersion, $\sigma_{v_{\rm
tot}}$, for all spectral types listed in Table~\ref{sigmas}.  Galactic
velocities comprise the three components of the space velocities and
as such, may summarize the statistical effects seen in each velocity
component (unless they are canceled out). The kinematical ages derived
from \citet{wielen77} equation are also provided in the Table. The
L--T population shows a kinematical age in the range 0.5--2.3\,Gyr; it
thus appears to be a factor of two younger than the low-mass stars of
the solar neighborhood. From the analysis of the tangential velocities
of a sample of late M and L dwarfs, \citet{dahn02} and
\citet{schmidt07} also concluded that these objects have a mean age of
2--4\,Gyr, consistent with our derivation. The fact that the late
M-type stars of the solar vicinity appear to be younger, on average,
than earlier-type stars was previously suggested in the literature
\citep{hawkins88,kirk94,reid94,reid02}. Here, we extend this result to
the coolest spectral types (L and T) including the brown dwarf
population.

Finally, various groups have also pointed out that nearby, very
low-mass stars tend to have smaller space motions compared to the
earlier-type stars, although the difference is claimed to be of
marginal significance \citep{reid02,dahn02}. We provide in
Table~\ref{sigmas} the average proper motion, $<\mu>$, for the F, G,
K, M0--M5, L and T dwarf populations with known trigonometric
parallaxes considered in this work (1-$\sigma$ dispersions are given
in brackets). Nevertheless, the information on the mean distance and
the mean proper motion is better encapsulated in the mean tangential
velocity, $<v_{\rm t}>$, which we also provide in
Table~\ref{sigmas}. To complete the spectral sampling between the
early M stars and the ultracool dwarfs of our study we have collected
the $UVW$ data of 29 M6--M9 objects at less than 23\,pc of the Sun
from Table~5 of \citet{reid02}. These authors obtained accurate radial
velocities for these objects and combined them with astrometric proper
motions and photometric parallaxes to compute Galactic velocities. We
show in Table~\ref{sigmas} our derivation of $\sigma_U, \sigma_V,
\sigma_W$, and the mean Galactic velocity, proper motion, and
tangential velocity for the reduced sample of M6--M9 objects extracted
from \citet{reid02}.

\begin{figure}
\epsscale{1.0}
\plottwo{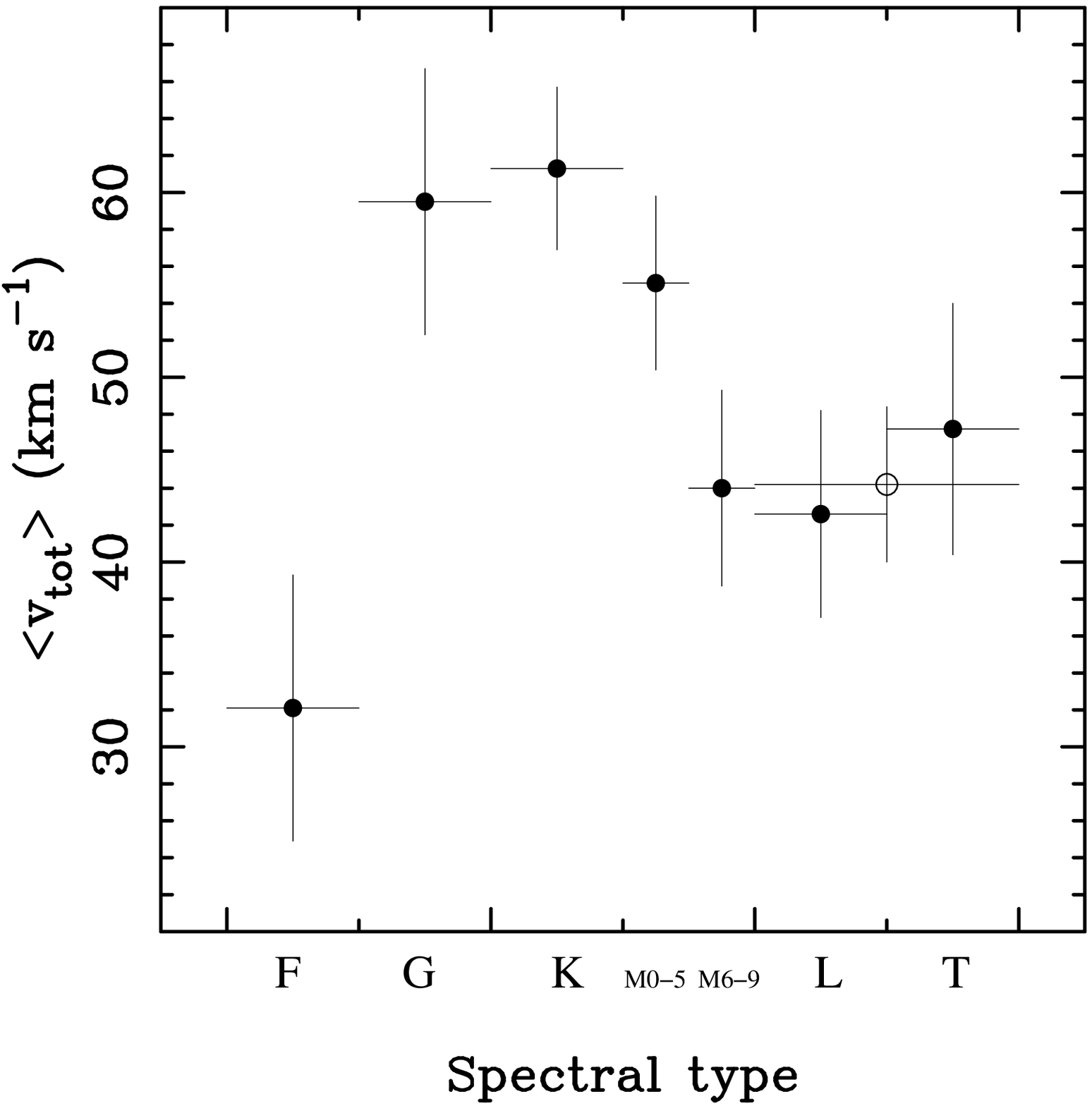}{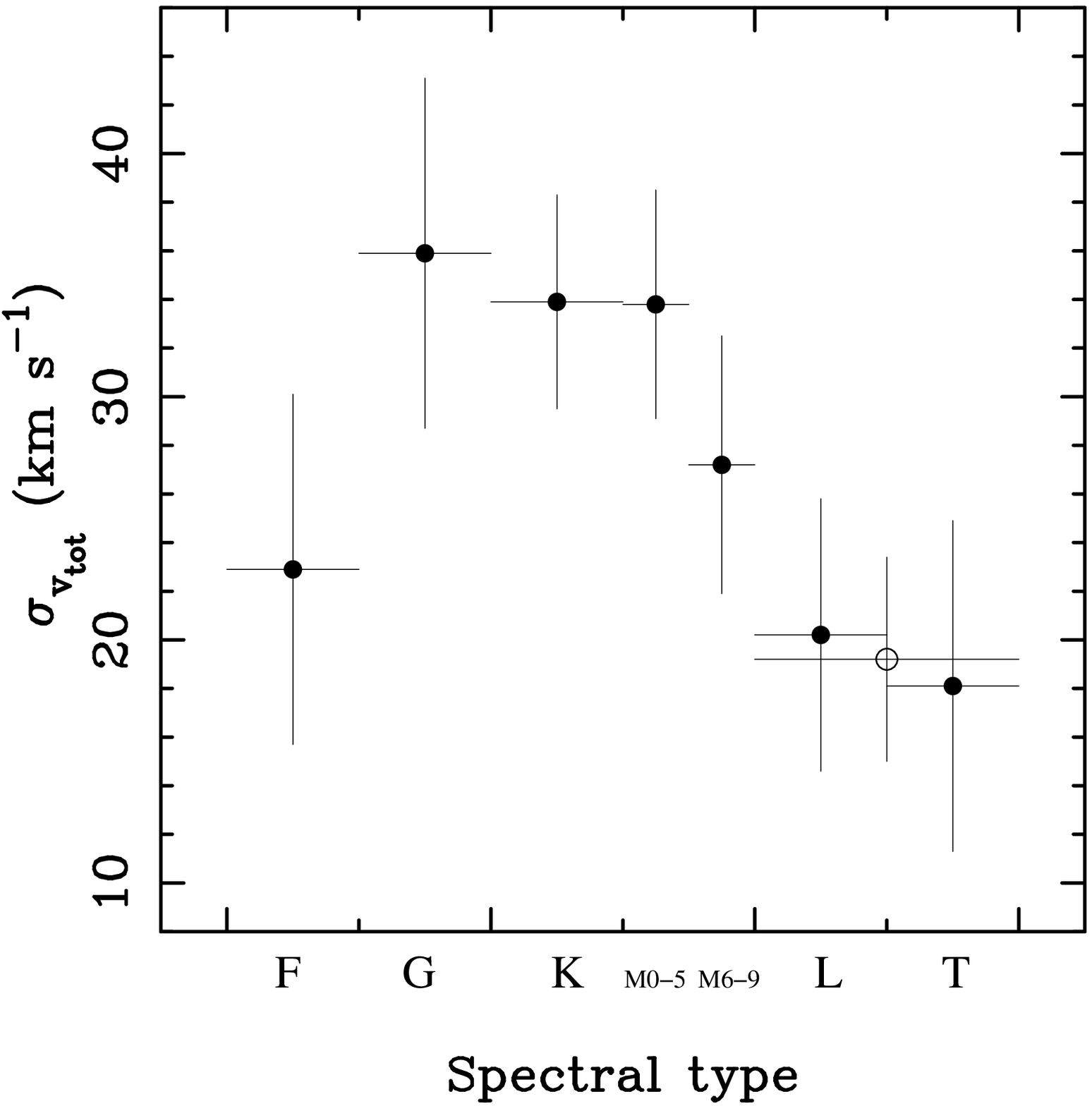}
\plottwo{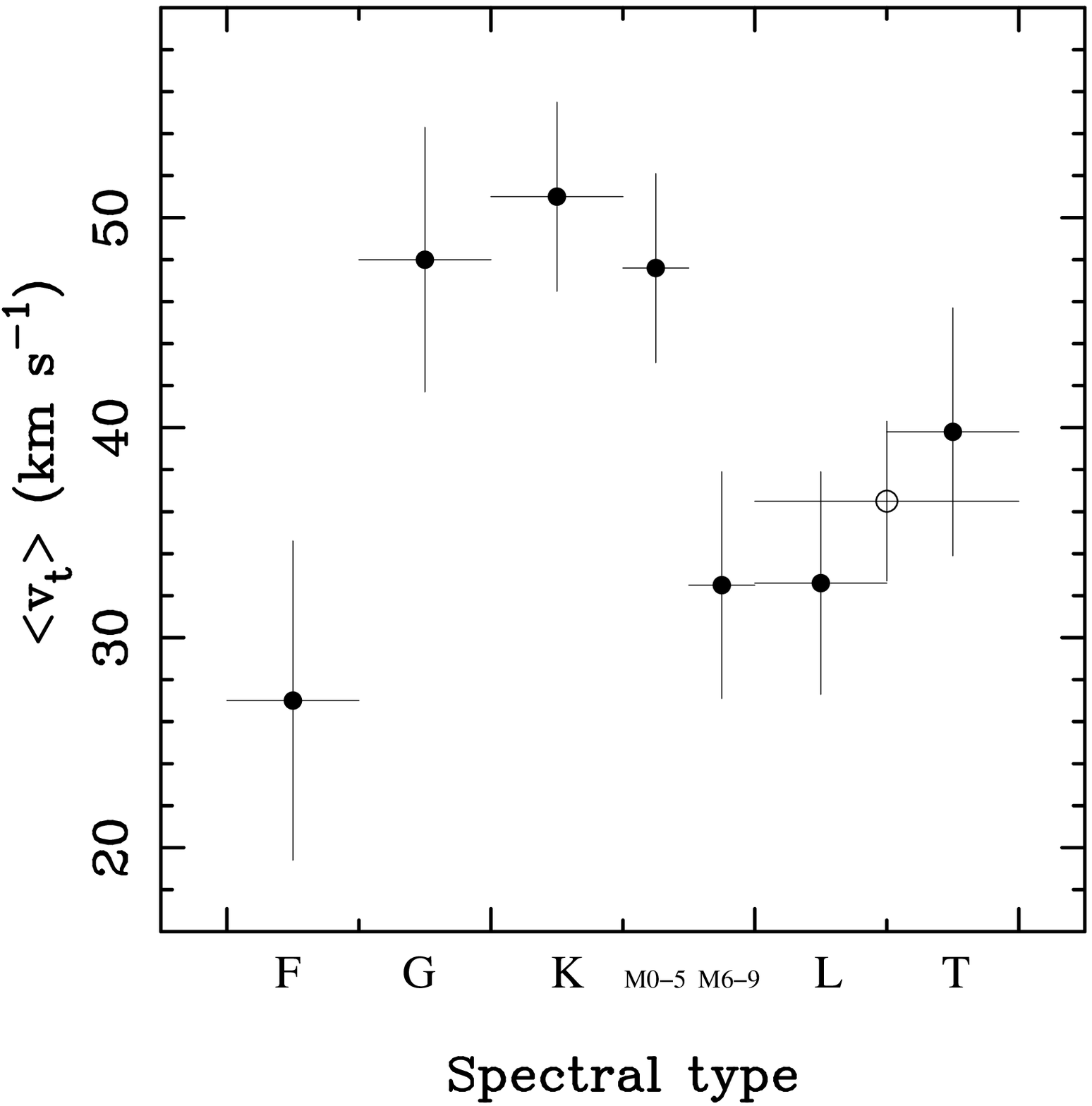}{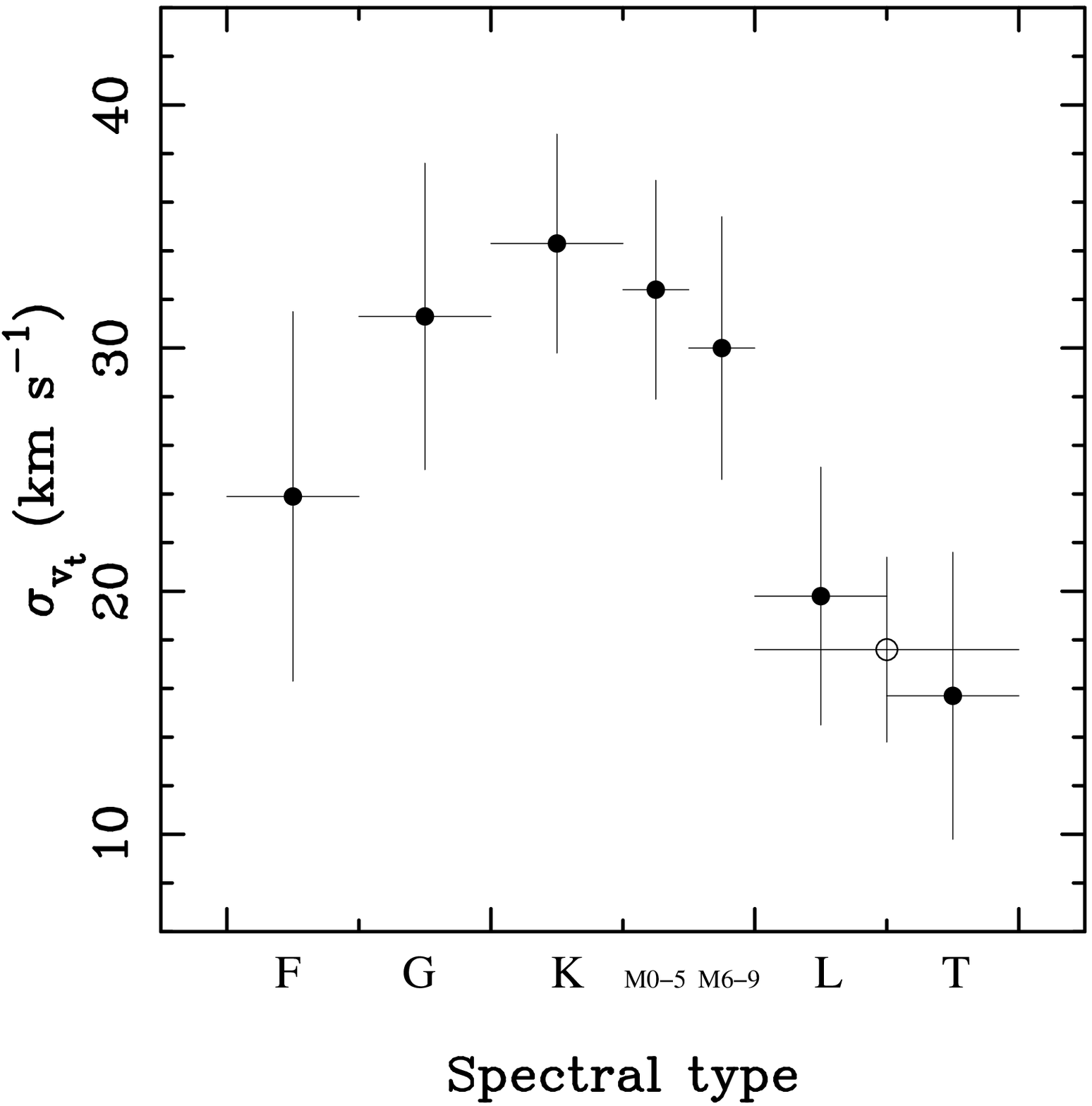}
\caption{{\sl (Left panels)} Average Galactic velocity ($<v_{\rm
tot}>$) and tangential velocity ($<v_{\rm t}>$) as a function of
spectral type. {\sl (Right panels)} Dispersion of Galactic
($\sigma_{v_{\rm tot}}$) and tangential ($\sigma_{v_{\rm t}}$)
velocities as a function of spectral type. The diagrams cover the mass
interval 2--0.05\,$M_\odot$. Vertical uncertainties correspond to the
error of the mean, while the horizontal error bars account for the
spectral range covered by individual measurements. All data are
volume-limited to $\sim$20\,pc. Note that we have split our sample of
ultracool dwarfs into the L and T classifications (filled
symbols). The open circle corresponds to the combined L and T
data. The values of F-type stars may need some correction (see text).
  \label{vgal_fig}}
\end{figure}

Figure~\ref{vgal_fig} illustrates the dependence of $<v_{\rm tot}>$,
$<v_{\rm t}>$ and their associated dispersions ($\sigma_{v_{\rm
tot}}$, $\sigma_{v_{\rm t}}$) on spectral type (or mass). We note that
the values corresponding to the F-type stars have less statistical
weight because they have been derived from a low number of
objects. After comparison with the literature values, the Galactic
velocity dispersion $\sigma_{v_{\rm tot}}$ of F stars shown in
Table~\ref{sigmas} is likely to be increased by a few \kms. The
diagrams of Fig.~\ref{vgal_fig} (which do not incorporate this
correction) cover a wide range of masses, from $\sim$2\,$M_{\odot}$
(F-type stars) down to $\sim$50\,\Mj~(or 0.05\,$M_{\odot}$,
T-types). Note that we have split our full sample of ultracool dwarfs
into the two L and T types. As seen in the right panels of
Fig.~\ref{vgal_fig}, K-type stars display the largest mean Galactic
and tangential velocities, while F stars and the ultracool dwarfs have
smaller values. We note that the mean velocities of M6--M9 stars are
comparable to those of the ultracool L and T dwarfs, in perfect
agreement with the results obtained for late-M and L-type objects by
\citet{schmidt07}.

The $\sigma_{v_{\rm tot}}$- and $\sigma_{v_{\rm t}}$-spectral type
relations shown in the right panels of Fig.~\ref{vgal_fig} present
similar structure: an increasing trend from F to K stars to decrease
at cooler types; the L and T population displays the smallest velocity
dispersions. We ascribe this behavior to the likely younger
kinematical ages of the ultracool and very low-mass objects in the
solar neighborhood. There is a hint in our data suggesting that T
dwarfs are slightly younger than L dwarfs; further observations and
larger samples are required to confirm it. This result may be affected
by an observational bias in the way L and T dwarfs are
discovered. Since brown dwarfs cool down and grow fainter with age, it
is expected that magnitude-limited surveys would find younger objects
in general. However, according to state-of-the-art evolutionary model
predictions on substellar temperatures and luminosities and the fact
that our study is limited to a distance of 20--23\,pc, this bias is
not largely contaminating our sample of ultracool dwarfs. Any model
explaining the population of the Galaxy should account for the
relations shown in Fig.~\ref{vgal_fig} quantitatively, which may
provide a constraint on the field mass function and the
stellar/substellar formation rate.

\section{Conclusions}

Radial velocities have been obtained for a sample of 18 M6.5--T8
dwarfs using high-resolution ($R$\,=\,17800--22700), near-infrared
(1.148--1.346\,$\mu$m) spectra collected with the NIRSPEC instrument
on the Keck\,II telescope. The sample comprises one M6.5 Pleiades
brown dwarf, two late-M field dwarfs, and 15 L and T likely brown
dwarfs of the solar neighborhood with masses in the range
30--75\,\Mj. Our radial velocity measurements further confirm the
membership of PPl\,1 in the Pleiades cluster, and the true physical
companionship of the T7.5V/T8V Gl\,570\,D dwarf in the Gl\,570
multiple system. From five velocity measurements obtained for
2MASS\,J05591914$-$1404488 (T4.5V) over 4.37\,yr and the observed
velocity dispersion of 0.5\,\kms, we discard the possible presence of
massive jovian planets near this brown dwarf with orbits of a few days
or around a year.

All the L and T dwarfs in our sample lie $\le$\,24\,pc from the Sun,
and trigonometric parallaxes and proper motions are available for
nearly all of them from \citet{dahn02}, \citet{vrba04}, and
\citet{knapp04}. We have used our radial velocity determinations and
astrometric data from the literature to derive Galactic $UVW$
velocities. The total number of ultracool dwarfs within 20\,pc of the
Sun is augmented up to 21 with the addition of eight L dwarfs of known
trigonometric distance, proper motion and radial velocity available in
the literature. We have compared the Galactic and tangential
velocities of the ultracool dwarf population to those of F, G, K, and
M stars contained in the same volumen. The $UVW$ distributions of the
complete expanded L and T sample show a smaller dispersion than the G
to M stars and a comparable dispersion to the F-type stars. Similar
behavior is observed for the total Galactic velocity and tangential
velocity dispersions, suggesting that the least massive population,
which includes brown dwarfs, is kinematically younger. Furthermore, we
find that a significant fraction ($\sim$40\%) of the L and T dwarfs in
our sample lies near the location of the Hyades moving group in the
$UV$ and $VW$ planes in contrast to the low rate (10--12\%) observed
for stars. This, in addition to the lower dispersions observed in all
space velocities, suggests that our sample of L and T dwarfs (many are
brown dwarfs) is kinematically young with likely ages below 5\,Gyr.

\acknowledgements We thank the referee for useful comments. The data
were obtained at the W. M. Keck Observatory, which is operated as a
scientific partnership between the California Institute of Technology,
the University of California, and NASA. The Observatory was made
possible by the generous financial support of the W. M. Keck
Foundation. The authors extend special thanks to those of Hawaiian
ancestry on whose sacred mountain we are privileged to be guests. We
thank the Keck observing assistants and the staff in Waimea for their
kind support. This research has made use of the SIMBAD database,
operated at CDS, Strasbourg, France, and has been supported by a Keck
PI Data Analysis grant awarded to E. L. M. by Michelson Science
Center, and by the Spanish projects AYA2003-05355 and
AYA2006-12612. We thank Terry Mahoney for his careful reading of the
English language. We also thank I$.$ Ribas for early discussions on
this topic.

\clearpage

\clearpage

\begin{deluxetable}{lcllrrrrrccc}
\tabletypesize{\scriptsize}
\rotate
\tablecaption{Heliocentric velocities of our sample. \label{vh}}
\tablewidth{0pt}
\tablehead{
\multicolumn{4}{c}{} & \multicolumn{5}{c}{Reference object} &  \multicolumn{1}{c}{} \\
\cline{5-9} \\
\colhead{Object}             & \colhead{SpT} & \colhead{Obs$.$ date} & \colhead{MJD} & \colhead{vB\,10\tablenotemark{a}} & \colhead{J2224$-$01\tablenotemark{a}} & \colhead{J0559$-$14\tablenotemark{a}} & \colhead{J1217$-$03\tablenotemark{a}} & \colhead{Other\tablenotemark{b}} & \colhead{$v_{\rm rot}$\,sin\,$i$\tablenotemark{c}} & Previous & Reference \\
\colhead{      }             & \colhead{   } & \colhead{   } & \colhead{($-$50000) } & \colhead{(km s$^{-1}$)        } & \colhead{(km s$^{-1}$)               } & \colhead{(km s$^{-1}$)               } & \colhead{(km s$^{-1}$)               } & \colhead{(km s$^{-1}$)} & \colhead{(km s$^{-1}$)} & \colhead{(km s$^{-1}$)} & \colhead{   } }
\startdata
PPl\,1                        & M6.5          & 2005 Oct 26  & 3669.51110 & $+$5.5$\pm$1.6 & \nodata             & \nodata         & \nodata         & \nodata         & 20.7    & $+$15.4$\pm$1.6 & 1 \\  
                              &               & 2005 Oct 28  & 3671.51447 & $+$5.3$\pm$1.9 & \nodata             & \nodata         & \nodata         &  $+$5.3$\pm$2.3 &         &         & \\  
vB10                          & M8V           & 2001 Jun 15  & 2075.58951 &$+$35.0$\pm$1.5\tablenotemark{d}&\nodata& \nodata       & \nodata         & \nodata & 6.5\tablenotemark{e}& $+$35.0 & \\  
                              &               & 2001 Nov 02  & 2215.20560 &$+$34.3$\pm$1.5\tablenotemark{d}&\nodata& \nodata       & \nodata         & \nodata         &         & $+$35.2$\pm$2.0 & 2 \\  
DENIS-P\,J033411.39$-$495333.6& M9V           & 2005 Oct 26  & 3669.48071 &$+$70.2$\pm$1.0 & \nodata             & \nodata         & \nodata         & \nodata         & $\le$15 &         & \\  
                              &               & 2005 Oct 28  & 3671.48317 &$+$69.8$\pm$1.0 & \nodata             & \nodata         & \nodata         & $+$69.8$\pm$1.0 &         &         & \\  
2MASS\,J00361617$+$1821104    & L3.5V/L4V     & 2004 Dec 05  & 3344.31265 &$+$18.2$\pm$1.6 & $+$18.6$\pm$1.4     & \nodata         & \nodata         & \nodata         & 36.0    &         & \\  
                              &               & 2005 Oct 28  & 3671.43118 & \nodata        & $+$17.6$\pm$1.5     & \nodata         & \nodata         & $+$17.7$\pm$1.0 &         &         & \\  
2MASS\,J22244381$-$0158521    & L4.5V         & 2001 Jun 15  & 2075.61089 &$-$37.8$\pm$1.0 & \nodata             & \nodata         & \nodata         & \nodata         & 30.7    & $-$37.4$\pm$3.4 & 3 \\  
SDSS\,J053951.99$-$005902.0   & L5V           & 2001 Nov 02  & 2215.45928 &$+$11.5$\pm$2.7 & $+$11.5$\pm$1.9     & \nodata         & \nodata         & \nodata         & 33.2    &         & \\  
                              &               & 2005 Oct 27  & 3670.50549 & \nodata        & $+$11.8$\pm$2.3     & \nodata         & \nodata         & $+$11.3$\pm$1.8 &         &         & \\  
2MASS\,J17281150$+$3948593AB  & L7V           & 2001 Jun 15  & 2075.54466 &$-$12.2$\pm$1.7 & $-$12.8$\pm$1.0     & \nodata         & \nodata         & \nodata         & 24.1    &         & \\  
2MASS\,J16322911$+$1904407    & L8V/L7.5V/dL6 & 2001 Jun 15  & 2075.51675 & $-$5.5$\pm$3.7 &  $-$5.9$\pm$2.2     & \nodata         & \nodata         & \nodata         & 21.8    & $+$5:   & 2 \\  
DENIS-P\,J0255.0$-$4700       & L9V/dL6     & 2005 Oct 27  & 3670.44084 &$+$17.9$\pm$3.6 & $+$17.5$\pm$2.8     & $+$16.0$\pm$3.8 & \nodata         & \nodata         & 41.0    & $+$13.0$\pm$2.0 & 2 \\  
SDSSp\,J125453.90$-$012247.4  & T2V           & 2006 Jan 19  & 3754.61049 & \nodata        &  $-$0.1$\pm$5.4     &  $+$0.9$\pm$2.2 & \nodata         & \nodata         & 28.4    &         & \\  
2MASS\,J05591914$-$1404488    & T4.5V         & 2001 Nov 02  & 2075.41290 &$-$13.1$\pm$2.8 & $-$13.8$\pm$3.7     & \nodata         & \nodata         & \nodata         & 22.8    &         & \\  
                              &               & 2004 Dec 05  & 3344.54854 & \nodata        & $-$14.6$\pm$1.3     & \nodata         & \nodata         & $-$14.0$\pm$3.8 &         &         & \\  
                              &               & 2005 Oct 26  & 3669.57615 & \nodata        & $-$13.7$\pm$1.6     & \nodata         & \nodata         & $-$13.6$\pm$1.6 &         &         & \\  
                              &               & 2005 Oct 27  & 3670.53068 & \nodata        & $-$14.2$\pm$1.6     & \nodata         & \nodata         & $-$13.0$\pm$1.9 &         &         & \\  
                              &               & 2005 Oct 28  & 3671.54024 & \nodata        & $-$14.3$\pm$1.9     & \nodata         & \nodata         & $-$13.8$\pm$1.0 &         &         & \\  
2MASS\,J15031961$+$2525196    & T5V           & 2006 Jan 19  & 3754.66827 & \nodata        & $-$40.1$\pm$5.8     & $-$39.9$\pm$2.6 & $-$40.5$\pm$2.1 & \nodata         & 32.1    &         & \\  
SDSS\,J162414.37$+$002915.6   & T6V           & 2001 Jun 15  & 2075.44304 & \nodata        & \nodata             & $-$30.7$\pm$3.0 & \nodata         & \nodata         & 36.6    &         & \\  
SDSS\,J134646.45$-$003150.4   & T6.5V         & 2001 Jun 15  & 2075.34518 & \nodata        & \nodata             & $-$23.1$\pm$1.5 & \nodata         & \nodata         & $\le$15 &         & \\  
2MASS\,J15530228$+$1532369    & T7V           & 2001 Jun 15  & 2075.41290 & \nodata        & \nodata             & $-$32.8$\pm$3.3 & $-$32.9$\pm$3.0 & \nodata         & 29.4    &         & \\  
2MASS\,J12171110$-$0311131    & T7V/T8V       & 2001 Jun 15  & 2075.25577 & \nodata        & \nodata             &  $+$5.0$\pm$1.6 & \nodata         & \nodata         & 31.4    &         & \\  
Gl\,570D                      & T7.5V/T8V     & 2001 Jun 15  & 2075.38494 & \nodata        & \nodata             & $+$28.9$\pm$2.4 & $+$28.2$\pm$3.1 & \nodata         & 30.7    &         & \\  
2MASS\,J04151954$-$0935066    & T8V           & 2005 Oct 26  & 3669.55271 & \nodata        & \nodata             & $+$49.6$\pm$1.2 & $+$51.0$\pm$4.3 & \nodata         & 33.5    &         & \\  
\enddata
\tablenotetext{a}{Heliocentric radial velocities (km\,s$^{-1}$) used in the cross-correlation: 
$-$35.0 (vB\,10), $-$37.8 (J2224$-$01), $-$13.8 (J0559$-$14), and $+$5.0 (J1217$-$03).}
\tablenotetext{b}{The dwarf's earliest spectrum acts as the reference spectrum in the cross-correlation.}
\tablenotetext{c}{Average rotational velocity from \citet{osorio06}.}
\tablenotetext{d}{Heliocentric velocity obtained from the centroids of the K\,{\sc i} lines.}
\tablenotetext{e}{Rotational velocity from \citet{mohanty03}.}
\tablerefs{(1) \citet{martin98}; (2) \citet{mohanty03}; (3) \citet{bailer04}.}
\end{deluxetable}


\begin{deluxetable}{lccrc}
\tablecaption{Mean heliocentric radial velocities. \label{meanv}}
\tablewidth{0pt}
\tablehead{
\colhead{Object} & \colhead{SpT} & \colhead{$\Delta$t} & \colhead{$<v_h>$} & \colhead{$N$} \\
\colhead{      } & \colhead{   } & \colhead{         } & \colhead{(\kms) } & \colhead{   } }
\startdata
PPl\,1     & M6.5      & 2.0 d   &  $+$5.4\,$\pm$\,1.7                  & 2 \\
J0334$-$49 & M9V       & 2.0 d   & $+$70.0\,$\pm$\,1.0                  & 2 \\
J0036$+$18 & L3.5V/L4V & 0.90 yr & $+$18.1\,$\pm$\,1.5                  & 2 \\
J0539$-$00 & L5V       & 3.98 yr & $+$11.7\,$\pm$\,2.1                  & 2 \\
J0559$-$14 & T4.5V     & 4.37 yr & $-$13.8\,$\pm$\,0.2\tablenotemark{a} & 5 \\
\enddata
\tablenotetext{a}{The associated velocity uncertainty corresponds to the error of the mean.}
\end{deluxetable}


\begin{deluxetable}{lcrrrrrr}
\tabletypesize{\scriptsize}
\tablecaption{Galactic velocities of our sample. \label{vgal}}
\tablewidth{0pt}
\tablehead{
\colhead{Object} & \colhead{SpT} & \colhead{$d$\tablenotemark{a}} & \colhead{$\mu_\alpha {\rm cos} \delta$\tablenotemark{a}} & \colhead{$\mu_\delta$\tablenotemark{a}} & \colhead{$U$}           & \colhead{$V$}           & \colhead{$W$}           \\
\colhead{      } & \colhead{      } & \colhead{(pc)}                 & \colhead{(mas\,yr$^{-1}$)             } & \colhead{(mas\,yr$^{-1}$)             } & \colhead{(km s$^{-1}$)} & \colhead{(km s$^{-1}$)} & \colhead{(km s$^{-1}$)} }
\startdata
PPl\,1\tablenotemark{b}     & M6.5V         & 133.8   &    $+$19.7 &   $-$44.8 &   $-$6.4$\pm$2.3  & $-$27.6$\pm$5.0  & $-$14.0$\pm$4.7  \\
vB\,10\tablenotemark{c}     & M8V           &   5.9   &   $-$578.8 & $-$1331.6 &  $+$52.6$\pm$0.8  &  $-$8.1$\pm$0.7  &  $-$5.1$\pm$0.2  \\
J0334$-$49\tablenotemark{d} & M9V           &   8.2   &  $+$2350.0 &  $+$470.0 &  $-$66.2$\pm$18.7 & $-$95.7$\pm$25.0 &  $-$5.8$\pm$21.7 \\
J0036$+$18                  & L3.5V/L4V     &   8.8   &   $+$899.1 &  $+$120.0 &  $-$40.0$\pm$0.8  &  $-$4.8$\pm$1.5  & $-$11.5$\pm$1.7  \\
J2224$-$01                  & L4.5V         &  11.5   &   $+$463.8 &  $-$865.8 &  $-$10.6$\pm$0.6  & $-$64.5$\pm$0.9  &  $-$6.4$\pm$0.8  \\
J0539$-$00                  & L5V           &  13.1   &   $+$164.4 &  $+$315.9 &  $-$20.0$\pm$1.8  &  $+$4.3$\pm$1.0  & $+$14.4$\pm$0.7  \\
J1728$+$39AB                & L7V           &  24.1   &    $+$25.9 &   $-$36.8 &   $-$0.4$\pm$1.0  &  $-$9.3$\pm$1.4  &  $-$9.8$\pm$1.1  \\
J1632$+$19                  & L8V/L7.5V/dL6 &  15.4   &   $+$294.5 &   $-$56.6 &   $+$5.3$\pm$1.4  &  $+$8.3$\pm$1.3  & $-$20.4$\pm$1.6  \\
J0255$-$47                  & L9V/dL6    &   6.3   &  $+$1036.6 &  $-$598.5 &   $-$5.4$\pm$5.0  & $-$39.2$\pm$5.1  &  $+$3.5$\pm$4.0  \\
J1254$-$01                  & T2V           &  11.8   &   $-$478.7 &  $+$130.1 &  $-$25.6$\pm$0.8  &  $-$9.7$\pm$0.9  &  $+$4.6$\pm$1.9  \\
J0559$-$14                  & T4.5V         &  10.3   &   $+$562.1 &  $-$342.4 &  $+$22.8$\pm$1.1  & $-$15.8$\pm$1.0  & $+$21.3$\pm$0.6  \\
J1624$+$00                  & T6V           &  11.2   &   $-$378.6 &    $-$2.5 &  $-$31.0$\pm$2.5  & $-$19.3$\pm$0.8  &  $-$2.3$\pm$1.7  \\
J1346$-$00                  & T6.5V         &  13.7   &   $-$473.1 &  $-$132.3 &  $-$29.3$\pm$1.8  & $-$20.0$\pm$1.6  & $-$17.4$\pm$1.4  \\
J1217$-$03                  & T7V/T8V       &   9.8   &  $-$1057.8 &   $+$91.6 &  $-$44.0$\pm$3.6  & $-$23.1$\pm$2.2  &  $-$0.2$\pm$1.5  \\
Gl\,570D\tablenotemark{c}   & T7.5V/T8V     &   5.9   &  $+$1034.1 & $-$1725.5 &  $+$49.6$\pm$2.0  & $-$22.3$\pm$0.8  & $-$31.6$\pm$1.4  \\
J0415$-$09                  & T8V           &   5.7   &  $+$2192.8 &  $+$527.3 &  $-$64.2$\pm$0.9  & $-$43.3$\pm$0.7  & $+$15.1$\pm$1.0  \\
\enddata
\tablenotetext{a}{From \citet{perryman97,dahn02,vrba04,knapp04,an07}.}
\tablenotetext{b}{Distance and proper motion of the Pleiades cluster.}
\tablenotetext{c}{Distance and proper motion of the primary stars \citep{perryman97}.}
\tablenotetext{d}{Spectroscopic parallax from \citet{phan-bao06}.}
\end{deluxetable}


\begin{deluxetable}{lcrrrcrrrl}
\tabletypesize{\scriptsize}
\tablecaption{Galactic velocities of an additional set of L-type dwarfs. \label{vgal_literature}}
\tablewidth{0pt}
\tablehead{
\colhead{Object} & \colhead{SpT} & \colhead{$d$} &
\colhead{$\mu_\alpha {\rm cos} \delta$} &
\colhead{$\mu_\delta$} & \colhead{$v_h$}         &
\colhead{$U$}           & \colhead{$V$}           & \colhead{$W$}           &
\colhead{Ref} \\
\colhead{      } & \colhead{      } & \colhead{(pc)}                 & \colhead{(mas\,yr$^{-1}$)             } & \colhead{(mas\,yr$^{-1}$)             } & \colhead{(km s$^{-1}$)} & \colhead{(km s$^{-1}$)} & \colhead{(km s$^{-1}$)} & \colhead{(km s$^{-1}$)} & \colhead{}  }
\startdata
2MASS\,J07464256$+$2000321 & L0.5V & 12.2 &  $-$374.044 &   $-$57.905 & $+$54  &  $-$56.2$\pm$4.4 &  $-$15.3$\pm$1.6 &   $-$0.7$\pm$1.8 &  1, 3 \\     
2MASS\,J14392836$+$1929149 & L1V   & 14.4 & $-$1229.791 &  $+$406.714 & $-$28  &  $-$82.6$\pm$2.1 &  $-$39.5$\pm$0.9 &  $+$13.2$\pm$4.5 &  1, 2, 3 \\  
Kelu\,1AB                  & L2V   & 18.7 &  $-$284.790 &   $+$10.941 & $+$17  &  $-$12.8$\pm$1.2 &  $-$24.6$\pm$1.4 &  $+$12.4$\pm$1.2 &  2, 3\\      
DENIS-P\,J1058.7$-$1548    & L3V   & 17.3 &  $-$252.931 &   $+$41.419 & $+$19  &  $-$20.3$\pm$0.3 &  $-$19.0$\pm$1.5 &   $+$5.5$\pm$1.3 &  2, 3\\      
DENIS-P\,J1228.2$-$1547AB  & L5V   & 20.2 &  $+$133.868 &  $-$179.598 &  $+$2.5&  $+$19.0$\pm$1.5 &   $-$5.4$\pm$3.1 &   $-$8.7$\pm$3.7 &  2, 3\\      
2MASS\,J15074769$-$1627386 & L5V   &  7.3 &  $-$161.476 &  $-$888.547 & $-$39  &  $-$25.3$\pm$3.9 &  $-$17.0$\pm$1.1 &  $-$39.8$\pm$2.9 &  1, 3\\      
DENIS-P\,J0205.4$-$1159AB  & L7V   & 19.8 &  $+$434.348 &   $+$54.871 &  $+$7  &  $-$34.8$\pm$2.1 &  $-$21.8$\pm$0.8 &   $+$6.4$\pm$4.6 &  2, 3\\      
2MASS\,J08251968$+$2115521 & L7.5V & 10.6 &  $-$506.522 &  $-$292.729 & $+$20  &  $-$27.3$\pm$4.0 &  $-$16.9$\pm$1.7 &  $-$15.1$\pm$2.5 &  1, 3, 4\\   
\enddata
\tablerefs{Radial velocities from: (1) \citet{bailer04}; (2) \citet{mohanty03}. Distances and proper motions from: (3) \citet{dahn02}; (4) \citet{vrba04}.}
\end{deluxetable}


\begin{deluxetable}{lcccccccccl}
\tabletypesize{\scriptsize}
\tablecaption{$UVW$, total Galactic and tangential velocity distributions of stars and brown dwarfs with $d$\,$\le$\,20~pc. \label{sigmas}}
\tablewidth{0pt}
\tablehead{
\colhead{SpT} & \colhead{$N$} & \colhead{$<d>$\tablenotemark{a}} & \colhead{$<\mu>$\tablenotemark{a}} & \colhead{$\sigma_U$} & \colhead{$\sigma_V$} & \colhead{$\sigma_W$} & \colhead{$<v_{\rm tot}>$\tablenotemark{a}} & \colhead{$<v_{\rm t}>$\tablenotemark{a}} & \colhead{Age} & \colhead{Ref.} \\
\colhead{} & \colhead{} & \colhead{(pc)} & \colhead{(\arcsec\,yr$^{-1}$)} & \colhead{(km\,s$^{-1}$)} & \colhead{(km\,s$^{-1}$)} & \colhead{(km\,s$^{-1}$)} & \colhead{(km\,s$^{-1}$)} & \colhead{(km\,s$^{-1}$)} & \colhead{(Gyr)} & \colhead{}
}
\startdata
F                        & 10 & 15.4 (3.4) & 0.44 (0.59) & 30.9 & 10.3 & 18.1 & 32.1 (22.9) & 27.0 (23.9) &  2.2$^{+3.0}_{-1.6}$ & 1, 5    \\
G                        & 25 & 15.6 (3.9) & 0.68 (0.82) & 41.3 & 29.3 & 20.5 & 59.5 (35.9) & 48.0 (31.3) &  9.0$^{+6.7}_{-4.5}$ & 1, 5    \\
K                        & 58 & 14.0 (3.4) & 0.78 (0.80) & 43.7 & 32.9 & 15.9 & 61.3 (33.9) & 51.0 (34.3) &  7.6$^{+3.4}_{-2.6}$ & 1, 5    \\
M0--M5                   & 52 & 10.3 (4.0) & 1.12 (1.38) & 47.1 & 30.7 & 19.5 & 55.1 (33.8) & 47.6 (32.4) &  7.5$^{+3.6}_{-2.7}$ & 1, 5    \\
M6--M9  & 31\tablenotemark{b} & 17.4 (5.9) & 0.45 (0.45) & 36.5 & 20.6 & 15.7 & 44.0 (29.6) & 32.5 (30.0) &  3.8$^{+2.8}_{-1.9}$ & 2, 8    \\
L and T\tablenotemark{c} & 13 & 10.3 (3.2) & 0.92 (0.64) & 30.0 & 19.9 & 15.2 & 43.8 (17.4) & 37.5 (14.3) &  0.9$^{+1.1}_{-0.6}$ & 2, 5--7 \\
L and T\tablenotemark{d} & 21 & 12.2 (4.4) & 0.77 (0.57) & 30.2 & 16.5 & 15.8 & 44.2 (19.1) & 36.5 (17.6) &  1.2$^{+1.1}_{-0.7}$ & 2--7    \\
\enddata
\tablenotetext{a}{ Values in brackets correspond to the width of the distributions (or $\sigma$).}
\tablenotetext{c}{ Includes the 29 objects extracted from \citet{reid02} and the two field M dwarfs in our sample.}
\tablenotetext{c}{ Sample shown in Table~\ref{vgal}.}
\tablenotetext{d}{ Extended sample.}
\tablerefs{Radial velocities from (1) \citet{kharchenko04}; (2) this paper; (3) \citet{mohanty03}; (4) \citet{bailer04}. Distances and proper motions from (5) \citet{perryman97}; (6) \citet{dahn02}; (7) \citet{vrba04}; (8) \citet{reid02}.}
\end{deluxetable}

\end{document}